\documentclass[11pt]{article}

\usepackage[utf8]{inputenc}
\usepackage[T1]{fontenc}
\usepackage{lmodern}
\usepackage[margin=1in]{geometry}
\usepackage{amsmath,amssymb}
\usepackage{graphicx}
\usepackage{siunitx}
\usepackage{booktabs}
\usepackage{authblk}
\usepackage[colorlinks=true,allcolors=blue]{hyperref}
\usepackage[capitalise,nameinlink]{cleveref}   

\graphicspath{{figs/}}
\crefname{figure}{Fig.}{Figs.}
\Crefname{figure}{Fig.}{Figs.}

\newcommand{\bmhead}[1]{\paragraph*{#1}}

\newcommand{\Eeff}{E_{\mathrm{eff}}}
\newcommand{\Eext}{E_{\mathrm{ext}}}

\newcommand{\dE}{{\mathrm{dE_{ce}}}}

\title{Resolving and resetting the charge environment of single T-centers in
silicon p--i--n waveguides}

\author[1,2,*]{Chaoshen Zhang}
\author[3,4]{Hanbin Song}
\author[3,5]{Lukasz Komza}
\author[1,2]{Aaron M. Day}
\author[3,6]{Enrique Garcia}
\author[2]{Donald Witt}
\author[7]{Mihir K. Bhaskar}
\author[3,5,6]{Alp Sipahigil}
\author[1,2]{Evelyn L. Hu}

\affil[1]{Harvard Quantum Initiative, Harvard University, Cambridge, MA 02138, USA}
\affil[2]{John A. Paulson School of Engineering and Applied Sciences, Harvard
University, Cambridge, MA 02138, USA}
\affil[3]{Materials Sciences Division, Lawrence Berkeley National Laboratory,
Berkeley, CA 94720, USA}
\affil[4]{Department of Materials Science and Engineering, University of
California, Berkeley, Berkeley, CA 94720, USA}
\affil[5]{Department of Physics, University of California, Berkeley, Berkeley,
CA 94720, USA}
\affil[6]{Department of Electrical Engineering and Computer Sciences, University
of California, Berkeley, Berkeley, CA 94720, USA}
\affil[7]{IonQ, 1284 Soldiers Field Rd, Boston, MA 02135, USA}
\affil[*]{Correspondence:
\href{mailto:chaoshen_zhang@g.harvard.edu}{chaoshen\_zhang@g.harvard.edu}}

\date{}

\begin{document}

\maketitle

\begin{abstract}
The silicon T-center is a telecom-band spin-photon interface in a manufacturable
photonics platform. In nanophotonic devices, however, its optical linewidth is
broadened by a fluctuating charge environment, limiting photon
indistinguishability for quantum networking. Here, we address this challenge
through characterization and suppression of the local charge-noise of single
T-centers in lateral p--i--n waveguides, demonstrating an unheralded method of
T-center optical linewidth narrowing. Above-band illumination resets the charge
environment, neutralizing the local field and suppressing spectral diffusion.
Across 46 emitters this reset narrows the median linewidth 3.5-fold to
0.57(11)~GHz, and an optimized emitter reaches 128(22)~MHz, the narrowest
unheralded linewidth reported for an integrated T-center. The narrowed
transition supports coherent optical Rabi oscillations with a coherence time of
20(2)~ns. Finally, we perform Stark-shift tuning using the p--i--n junction, and
read out the local electric field and the charge-noise width at 15~mK. An
analytical model of proximal surfaces, bulk, and junction field effects provides
good agreement with our findings. Our multi-pronged characterization of the
nanophotonic-integrated T-center charge environment enables future device
optimization toward scalable quantum interconnects.
\end{abstract}

\section{Introduction}

Solid-state spin--photon interfaces (SPIs) are elementary
building blocks of quantum networks and distributed quantum computing~\cite{awschalom2018spins}.
SPIs in wide-bandgap hosts such as diamond, integrated into nanophotonic devices, have demonstrated the essential primitives, from spin--photon entanglement to multi-qubit registers~\cite{stas2022node,knaut2024entanglement}.
However, realizing them at scale favors a candidate that radiates in the low-loss
telecommunication band and is native to a manufacturable platform.
The silicon T-center satisfies these requirements: it emits in the telecom
O-band near 1326~nm~\cite{bergeron2020tcenter}, can be created directly in
silicon-on-insulator (SOI), inherits the mature silicon photonics
manufacturing technology~\cite{johnston2024cavity,komza2025multiplexed}, and
hosts an addressable electron spin with a long-lived nuclear-spin register~\cite{higginbottom2022singlespin,song2026entanglement}.
An open challenge for semiconductor SPIs is demonstrating
sufficient optical coherence for quantum information schemes~\cite{janitz2020,Borregaard2020},
alongside the spectral tuning needed to overcome their inhomogeneous distribution.

The T-center's promise is presently inhibited by exactly this challenge.
For T-centers in nanophotonic devices, the resonant photoluminescence-excitation (PLE)
linewidth of a single T-center is typically a few GHz~\cite{higginbottom2022singlespin}, four orders
of magnitude above the 0.17~MHz lifetime limit and far broader than the 33~MHz
ensemble linewidth measured in bulk isotopically enriched silicon~\cite{bergeron2020tcenter}.
Recent work attributes this gap to spectral diffusion driven by a
resonant-laser-pumped charge environment around the emitter.
Heralding protocols recover emitter linewidths below 200~MHz, but only by
conditional selection that reduces system efficiency~\cite{zhang2025spectraldiffusion,bowness2025spectral}.
The T-center is especially exposed to this noise: while its linear Stark
susceptibility (a permanent dipole moment change of $\sim$1~D) is comparable to
its counterparts in diamond and 4H-silicon carbide (SiC)~\cite{tamarat2006stark,delascasas2017divacancy}, its quadratic polarizability
is at least an order of magnitude larger, 0.115~Hz\,m$^{2}$/V$^{2}$~(\cite{clear2024transition}, Methods), a consequence of its bound-exciton nature~\cite{alaerts2025starkshift}.
However, this same sensitivity to charge fluctuations renders an opportunity for
electrical tuning and control.
A scalable silicon SPI therefore needs two capabilities at once: control of the charge
environment that sets the linewidth, and electrical tuning of the zero-phonon
line (ZPL) to bring emitters into mutual resonance.
Emitter integration into p--i--n diodes has proven a powerful tool for
engineering coherence in other platforms~\cite{anderson2019electrical}, yet
electrical control of T-centers has only begun to be explored, with recent
demonstrations of Stark tuning~\cite{clear2024transition,dobinson2026spectraltuning} and electroluminescence~\cite{dobinson2025electrical,day2025ndr}.
Critically, unheralded narrowing of the T-center optical linewidth remains to
be demonstrated.

In this work we integrate individual T-centers into a 7~$\mu$m-wide p--i--n waveguide
and establish electrical and optical control over their charge environment.
Calibrating the spectral response against an externally applied field resolves
how each emitter shifts under bias.
We develop a model that converts that response into the external field at the emitter and the width of the charge-noise distribution, which reports on local variations in defect occupancy and sets its spectral diffusion.
Optically, an above-band (AB) illumination pulse neutralizes the steady-state
field and repopulates the charge traps surrounding the T-center; we optimize the
reset pulse and resolve its microsecond recovery dynamics.
Across 46 spatially and spectrally isolated emitters the reset reduces the median PLE
linewidth by 3.5$\times$, with selected centers reaching 128(22)~MHz.
Referencing each emitter's AB-PLE center frequency to its near-neutral frequency,
we extract the distribution of equilibrated fields and a charge-noise width
converging on a median value of $\sigma_E = 25.3$~kV/m, from which we propose and
weigh the relative contributions of two geometry-related broadening mechanisms.
The narrowed line supports coherent optical Rabi oscillations, letting us probe
the transition coherence across temperature with no significant variation in the linewidth.
Together these results establish electrical and optical control of the T-center
charge environment and point to the material optimization that a more coherent,
stable telecom emitter in silicon will require.

\section{Single T-centers in a p--i--n waveguide}

\begin{figure}[htbp]
    \centering
    \includegraphics[width=1\linewidth]{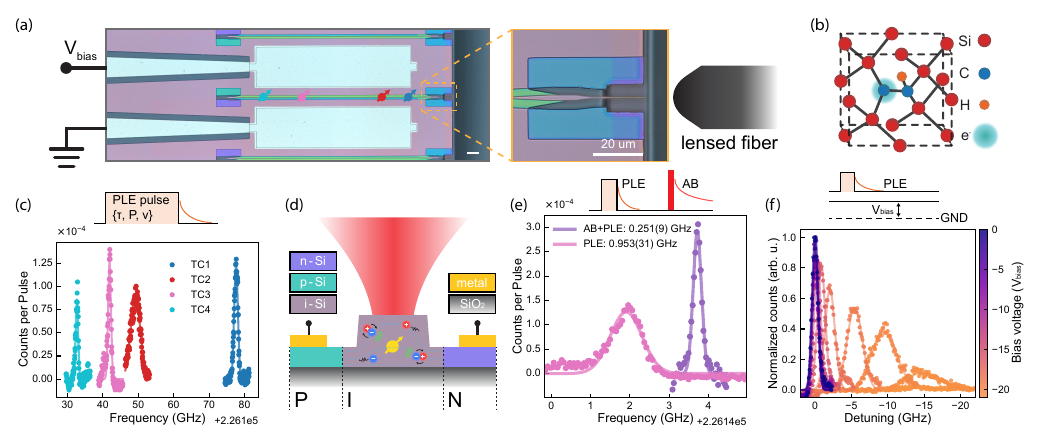}
    \caption{\textbf{Electrical and optical control of individual
    T-centers} (a) Optical micrograph of the device (true color).
    The partially etched ridge waveguide crosses the 7~$\mu$m intrinsic gap of a
    lateral p--i--n junction, giving the biased electrodes continuous field
    access to emitters along the device. Colored markers indicate the four
    T-centers (TC1--4) studied here.
    Photons are collected through a fully etched and released inverse taper at
    the right-hand facet (inset) into a lensed single-mode fiber, with a Bragg
    grating at the left end back-reflecting into the same path.
    Scale bars: 20~$\mu$m.
    (b) Atomic structure of a T-center.
    (c) Pulse sequence for PLE measurement; the exponential tail indicates the
    collection window (top).
    Background-subtracted PLE spectra of four T-centers (TC1--4), with linewidth of
    0.8--3.1~GHz, spectral-diffusion-dominated (bottom).
    (d) Device cross-section showing the T-center interacting with field,
    charge environment and free-space AB illumination.
    (e) Pulse sequence for AB-PLE, with a 6~$\mu$s AB-probe delay chosen to
    optimize SNR against the elevated AB background (top).
    Background-subtracted AB-PLE and PLE spectra on TC3: AB-PLE is
    3.8$\times$ narrower and 1.81~GHz blue-shifted (bottom).
    (f) Pulse sequence for DC-biased PLE measurement (top).
    TC3 DC-biased PLE spectra (bottom): the ZPL is insensitive to bias from 0 to $-15$~V, then shifts steeply beyond it.}
    \label{fig:intro}
\end{figure}

We study single T-centers implanted into the intrinsic region of a lateral
p--i--n junction, patterned in a 220~nm silicon-on-insulator (SOI) ridge
waveguide (\cref{fig:intro}a; Methods).
The junction width is varied from 3 to 14~$\mu$m.
The T-center is a carbon--hydrogen point defect with two carbon atoms sharing
one silicon site, a bound hydrogen, and an unpaired electron localized on the
other carbon (\cref{fig:intro}b)~\cite{dhaliah2022tcenter}.
Its formation yield is low, with reported optically detected T-center densities corresponding to yields on the order of (10$^{-5}$--10$^{-3}$) per implanted carbon ion~\cite{song2026entanglement}, 1--3 orders of magnitude below those of other solid-state emitters~\cite{pezzagna2010creation,schroder2017scalable,pavunny2021arrays}. Consequently, residual implanted dopants and lattice damage remain near each center and contribute to the fluctuating charge environment around each T-center~\cite{day2025ndr}.

Unless otherwise noted, emitters are addressed by free-space resonant excitation
with lensed single-mode fiber collection at a mixing chamber temperature of 15~mK
(Methods).
Two controls act at the emitter (\cref{fig:intro}d): a junction bias that sets a
controllable electric field (DC or pulsed), and local above-band (AB)
illumination from a co-propagating 640~nm pulsed laser that resets the charge
environment.
Carbon implantation density is set to $1\times10^{12}$~cm$^{-2}$ so roughly 15 spectrally and spatially resolvable T-centers are detected along a 300~$\mu$m waveguide (Methods).
Three resonant measurement regimes were used: PLE, the baseline scan; AB-PLE, a
PLE scan preceded by an above-band reset pulse; and DC-biased PLE, a PLE scan
under junction biasing (\cref{fig:intro}c,e,f).
In a representative waveguide with a 7~$\mu$m junction,
we focus on four centers, TC1--4, chosen for count rates that shorten integration times (\cref{fig:intro}c).
At 15~mK their PLE linewidths (FWHM) are 0.8--3.1~GHz, a two-orders-of-magnitude
broadening in contrast to the 33~MHz ensemble line of bulk isotopically
enriched silicon at 1.4~K~\cite{bergeron2020tcenter}.

Above-band illumination markedly reshapes the spectral response (\cref{fig:intro}e).
Following an unattenuated AB pulse (39~ns, 0.33~mW, 640~nm), the resonant probe is applied
after a 6~$\mu$s delay, and the AB-PLE spectrum overlays background-defect emission.
For TC3, AB-PLE is 3.8~$\times$ narrower and ZPL is blue-shifted by 1.81~GHz relative to
PLE center frequency.
We attribute these variations to the optical reset suppressing the charge noise
and canceling the field shift, as analyzed in detail in the following sections
(\crefrange{fig:field}{fig:dynamics}).
Under DC bias (\cref{fig:intro}f) the PLE line red-shifts and broadens as the magnitude
increases following a second-order Stark response~\cite{dobinson2026spectraltuning}.
This indicates that DC Stark tuning alone is an unfavorable route to
multi-emitter spectral alignment, motivating the field and reset studies that
follow.
\section{Electric-field tuning and charge environment mapping}

Having observed the impact of AB-illumination as optical reset, we now study the underlying broadening mechanism by further characterizing the field
experienced by the T-center.
We can understand the picture in two frames: the field at equilibrium, and that
of the environment represented by stochastic Gaussian noise.
Using the PIN-junction as a tool, we can independently probe each contribution.
A controlled bias shifts the ZPL through the Stark effect and broadens it through charge noise; the shift calibrates the emitter as a sensor of the local field, and the width correlates to the field fluctuations.

To vary the field while keeping the charge environment stable, we drive the
junction with an arbitrary-waveform generator (AWG): the square wave high level
($\mathrm{V_{bias}}$) overlaps the PLE probe, and low level (0~V) is paired
with an off-resonant stabilization pulse of comparable power and
$\sim$10$\times$ longer duration (Methods).
The IR stabilization pulse reinitializes the charge environment to a reproducible configuration between probes.
It counters the light-induced space-charge rearrangement that otherwise accentuates at
high field, weakening the screening and destabilizing the emitter frequency~\cite{zhou1998photorefractive,moia2022screening}.

\begin{figure}[htbp]
    \centering
    \includegraphics[width=1\linewidth]{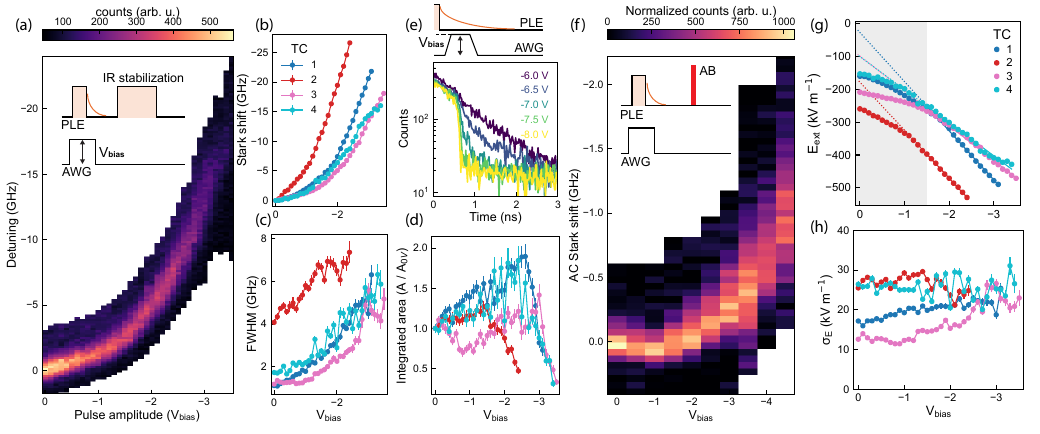}
    \caption{\textbf{Electric-field tuning and charge-noise readout
    with pulsed biasing} In (a)--(e) the junction is driven by a square wave
    with an off-resonant (1325~nm) IR stabilization pulse at 0~V; (f) replaces
    the IR stabilization pulse with an (640~nm) AB reset pulse.
    (a) AWG-bias pulse sequence (inset) and voltage-resolved PLE spectrogram of
    TC3: the ZPL shifts parabolically and broadens with $|V|$.
    (b) Relative Stark shift for TC1--4.
    (c) Gaussian-fit linewidth for TC1--4, steadily increasing with $|V|$.
    (d) Normalized integrated area for TC1--4: all show initial rise (with TC1
    approaching 1.9 times original area near $-2.5$~V) followed by a sharp
    decrease at high bias (ionization).
    (e) Ramped-bias lifetime of TC2, quenching near $-7.5$~V.
    (f) AB-PLE bias resolved spectrogram of TC3; the junction bias to external
    field gain is 4.1$\times$ smaller than in (a) due to screening.
    (g) AB-calibrated external field $\Eext$ versus junction bias, showing
    linear response $1\text{-}1.5 \times10^5~\mathrm{m^{-1}}$ at high bias
    ($<-1.5$~V).
    (h) r.m.s. charge-noise field $\sigma_E$ versus bias, TC2\&4 show no obvious
    trend, while $\sigma_E$ of TC1\&3 moderately increase with junction bias.
    Average $\sigma_E$ at high bias is 22.7~kV/m across the sweep.}
    \label{fig:field}
\end{figure}

\cref{fig:field}a shows the voltage-resolved PLE spectrogram of TC3.
Across TC1--4 the ZPL shifts parabolically with bias (\cref{fig:field}b), tracing the
second-order Stark response to the external field $\Eext$ at the emitter.
The emitter frequency exhibits a >25~GHz tuning range, with all four emitters
shifting over 15~GHz with less than 3.2~V of applied bias.
The FWHM grows monotonically with $|V|$ (\cref{fig:field}c).
The integrated intensity first rises for all four emitters, and by nearly twofold
for TC1, signaling an increased radiative-decay coupling into the waveguide which might originate from field-induced alignment of the excited-state dipole with the TE mode. This increase is followed by a drop above a threshold bias (\cref{fig:field}d) which we attribute to field
ionization of the T-center at high transient voltage~\cite{ganichev2000poolefrenkel}.
In a ramped-bias measurement, the TC2 fluorescence decay is unperturbed at low voltage but
is abruptly quenched once the ramp reaches $-7.5$~V (\cref{fig:field}e), limiting the
spectral tuning range.
The transient dynamics of an unramped square wave could ionize at lower nominal
bias, and the resonant illumination might also have lowered the ionization
threshold.

Replacing the IR stabilization pulse with an above-band (AB) reset pulse
significantly changes the response (\cref{fig:field}f).
Photocarriers fill the charge traps around the emitter to neutralize the equilibrated
field set up by device geometry, lowering the charge-configuration entropy and
narrowing the statistical distribution.
Correspondingly, the ZPL is blue-shifted from its PLE frequency.
With the transient injection of photocarriers, refilled surface and shallow bulk traps screen
much of the applied field, and thus the junction bias to external field gain
($\Eext/V_{\mathrm{bias}}$) is 4.1$\times$ smaller than AWG-pulse PLE.
Even with AB-illumination, we find the linewidth maintains a $|V|$ dependence.

An analytical model ties these observations together.
The net field at the emitter is $\Eeff = \Eext + \mathcal{N}(0,\sigma_E)$: an
equilibrated systematic part $\Eext$ ($\mathrm{V_{bias}}$ $+$ steady state
band-bending) and a Gaussian charge-noise profile of width $\sigma_E$, both
mapped to frequency by the quadratic Stark response $\Delta\nu = A\Eeff +
\alpha\Eeff^{2}$~\cite{clear2024transition}.
Taking the AB reset spectrum which we measured in a near-neutral charge
environment as the zero-field calibration, we use the Stark map to infer $\Eext$, which is linear in the applied bias.
The same map converts the linewidth into $\sigma_E$, which carries no
systematic dependence on the applied field: the linewidth grows with $|V|$ not
because the noise grows, but because the local Stark slope $g = A +
2\alpha\Eext$ steepens along the parabola, transducing a fixed $\sigma_E$ into a
wider frequency spread.

DC biasing produces similar Stark tuning range; however, it is strongly
nonlinear in turn-on behavior, varies from emitter to emitter
~\cite{clear2024transition,anderson2019electrical,day2024electrical}, and does not bring emitters
cleanly into mutual resonance with its long term frequency drift.
Such variation in tunability and spectral stability likely reflects differences
in T-center orientation, position in the waveguide cross-section, and local
charge environment. In-situ reverse bias IV measurement suggests that the leakage
current plays a minimal role in heating the junction and causing spectral drift.

\section{Optical reset of the charge environment}

We have established above that the PIN can control the equilibrated-static
field.
However, the second part of the charge environment--stochastic charge
noise--cannot be reduced by PIN biasing.
While AB-illumination is capable of reducing linewidth (\cref{fig:intro}), its exact
mechanism and influence on field noise has not been explored.
Next, we study the impact of the pulse control parameters and
resultant field dynamics.

\begin{figure}[htbp]
    \centering
    \includegraphics[width=0.6\linewidth]{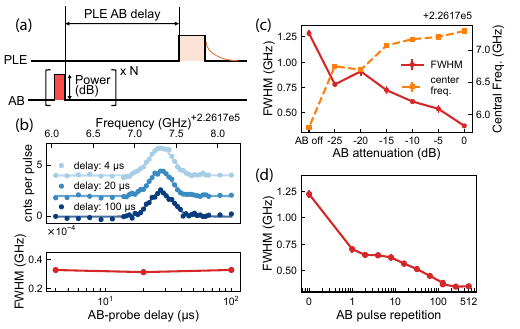}
    \caption{\textbf{Above-band pulse-parameter dependence} (a) AB-PLE
    pulse sequence, real time span not to scale.
    (b) TC1 AB-PLE spectra versus AB-probe delay (background subtracted); the
    linewidth is flat from 4 to 100~$\mu$s with a maximum variation of
    15(17)~MHz, no meaningful center wavelength drift is measured.
    (c) TC1 AB-PLE center wavelength and linewidth drifting toward the PLE
    (AB-off) values as the AB power is attenuated.
    (d) TC1 AB-PLE linewidth narrowing with AB-pulse repetition, saturating near
    350(13)~MHz at 256 pulses (25~dB attenuation).}
    \label{fig:absweep}
\end{figure}

We first map the reset effect against three AB-pulse parameters (\cref{fig:absweep}a): the
AB-probe delay, the reset pulse power, and the number of pulses.
We find that the line is insensitive to the delay from 4 to 100~$\mu$s with a
fixed FWHM and center frequency (\cref{fig:absweep}b).
We hence conclude that the reset charge configuration is stable in the dark for up to 100~$\mu$s,
consistent with the check-probe scheme~\cite{zhang2025spectraldiffusion}.
The stable center wavelength confirms that the
neutralized charge environment is not due to transient carrier-related effects.
Lowering the AB power, the line broadens and red-shifts toward its inherent PLE
position (\cref{fig:absweep}c): the AB pulse and the resonant probe compete, the former
filling traps toward neutrality while the latter repopulates and polarizes the
overall charge environment~\cite{ganichev2000poolefrenkel}.
Finally, increasing the number of (attenuated) AB pulses narrows the line by
60~\% to a saturated 350(13)~MHz floor by 256 pulses (\cref{fig:absweep}d).

\begin{figure}[htbp]
    \centering
    \includegraphics[width=0.6\linewidth]{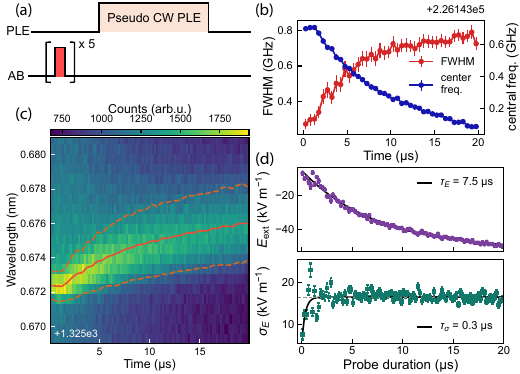}
    \caption{\textbf{Time-resolved charge-environment recovery} (a)
    Pseudo-CW-PLE sequence: five AB pulses reset the environment, then a
    20~$\mu$s resonant pulse probes while photons are registered (SNR $>$5 for
    TC3).
    (b) Fitted center frequency and linewidth versus probe time after reset.
    (c) Time-resolved spectrogram of TC3 under 100~$\mu$W resonant illumination.
    (d) Model-extracted recovery of the external field $\Eext(t)$ and the
    charge-noise amplitude $\sigma_E(t)$, on markedly different timescales.}
    \label{fig:dynamics}
\end{figure}

With this narrowing scheme characterized, we resolve the dynamics of the
charge environment recovery.
After five AB pulses, we collect fluorescence during a 20~$\mu$s resonant pulse (pseudo-CW
PLE; Methods) and track the spectra evolution (\cref{fig:dynamics}a--c).
Calibrating the field to the center wavelength under AB-illumination, the field
rebuilds from near neutrality toward its steady-state value ($-52$~kV/m for TC3,
$\tau = 7.5(2)$~$\mu$s) as the resonant probe re-polarizes the charge
environment.
The microsecond-order recovery reaches only about half the eventual equilibrated
field, so slower dynamics beyond 20~$\mu$s are not captured here.

\begin{figure}[htbp]
    \centering
    \includegraphics[width=0.6\linewidth]{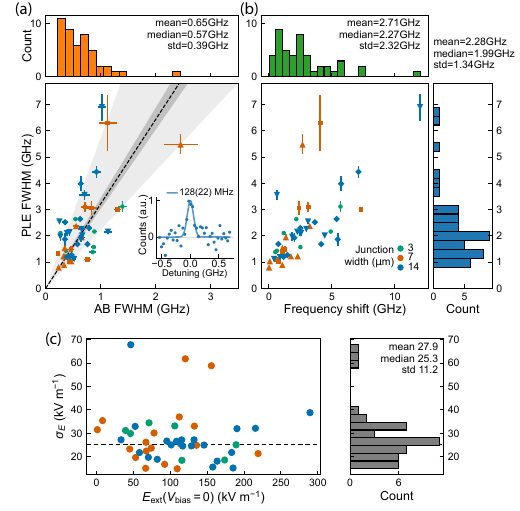}
    \caption{\textbf{Emitters statistics of the reset} (a) AB-PLE
    linewidth histogram over 46 isolated T-centers (top) and AB-PLE versus PLE
    linewidth correlation (bottom), data taken across 5 devices with 3 junction
    width (color-coded).
    Median linewidth reduced from 1.99 to 0.57~GHz, a 71~\% reduction.
    20~$\mu$W, 50~ns pulse on a selected emitter yields 128(22)~MHz fitted linewidth
    (inset).
    (b) AB-to-PLE frequency-shift histogram (top) and its correlation with PLE
    linewidth (bottom); PLE linewidth histogram (right).
    (c) Model derived $\Eext$ versus $\sigma_E$ correlation (left), $\sigma_E$
    histogram.
    Correlation shows -0.9~kV/m in $\sigma_E$ per 100~kV/m $\Eext$ increase,
    with median $\sigma_E$ of 25.3~kV/m.}
    \label{fig:stats}
\end{figure}

Finally, an automated screen spectrally locates and spatially identifies single emitters
along each waveguide and records a matched PLE / AB-PLE pair for each, across five devices (\cref{fig:stats}a; Methods).
For 46 individual emitters, AB-PLE narrows the line by
a median factor of 3.5, with the narrowest linewidth of selected emitter
reaching 128(22)~MHz--a record narrow T-center linewidth in an integrated device
without heralding~(\cref{fig:stats}a); the AB-to-PLE blue-shift increases with PLE linewidth
(\cref{fig:stats}b), both signatures of the magnitude of $\Eext$.
Converting each emitter's shift into its PLE-induced equilibrated field, the
linewidth returns a median charge-noise spread width $\sigma_E = 25.3$~kV/m
(16--84\% range 18.7--33.1~kV/m), independent of the field (\cref{fig:stats}c) and
consistent with the bias-series result (\cref{fig:field}).
The screen necessarily misses emitters too broad (low SNR) or too narrow (missed
in spatial scan), biasing against the extremes of $\sigma_E$.
Additionally, simplified scalar values were used and the absolute scale carries
a calibration systematic from the bulk Stark coefficient~\cite{clear2024transition} and the PLE-vs-AB reference (Methods).

The static and fluctuating parts of the field at the emitter have two candidate
sources: residual bulk traps and traps at the silicon interfaces.
The two types of traps have very different geometrical distributions.
Bulk traps are distributed randomly around the T-center; interface traps form
two sheets, the Si--vacuum surface above and the Si--SiO$_2$ buried-oxide (BOX)
boundary below, separated by the 220~nm device layer with the emitter near its
midplane.
We assess how each contributes to the two components of the field, $\Eext$ and
$\sigma_E$, and how transient photocarrier injection changes that contribution.

Bulk traps comprise levels native to the high-resistivity silicon, thermal
donors formed during the activation anneal, and the implantation damage that
accompanies T-center formation~\cite{kamiura1995deepcenter,sugiyama2012carbon}.
The deep levels among them contribute to the emitter's fluctuating environment:
under the junction field and below-band illumination they capture and release
charge, imposing a stochastic field on the T-center that registers as
$\sigma_E$.
They add little to $\Eext$, however, because their random positions and occupancies leave
the vector sum of their fields with a large variance but no preferred direction.
Photogenerated carriers in the intrinsic region drift to the p- and n-doped
regions under the built-in field, leaving the trap population near the emitter
in a steady state whose occupancy fluctuates under probe pulsing without
accumulating net charge.
Any static term that survives is therefore mediated by the junction
band-bending, and so by the junction width.
However, we find no significant dependence of $\Eext$ on that width (\cref{fig:stats}c),
so bulk traps are unlikely to dominate the equilibrated field.

Interface traps sit at the Si--vacuum and Si--BOX boundaries.
They can be comparably abundant, and present a continuum of levels at the band edge~\cite{poindexter1984pbcenters,lee1999etchdamage},
whose fluctuating occupancy feeds $\sigma_E$ alongside the bulk.
Unlike the bulk, an interface also bends the bands with its space-charge region, and the asymmetry between a
Si--vacuum surface and a Si--SiO$_2$ boundary leaves a net band-bending that
shifts the T-center's transition frequency.
Below-band illumination may accentuate such asymmetry by driving further charge accumulation at the
interfaces~\cite{kronik1999photovoltage}.
We therefore expect the interfaces rather than the bulk to set the static field
at the emitter.

Above-band injection acts on both.
The transient photocarriers neutralize the static field; recombination at the
nearby interfaces then removes them, freezing that neutral configuration as the
material reverts to an insulating behavior.
The AB-illuminated bias series is consistent with this: with the reset in place,
the field induced per applied volt falls by 4.1$\times$ (\cref{fig:field}f), as expected
once refilled surface and shallow traps screen the external electric field.
The accompanying drop in $\sigma_E$ we attribute mainly to the interfaces: as
the occupancy at both sheets rises, fewer states remain free to switch and the
noise falls.
The recovery dynamics emphasize the same distinction (\cref{fig:dynamics}): $\sigma_E$ has
largely returned within 0.3~$\mu$s, while $\Eext$ rebuilds twenty times more
slowly, over 7.5(2)~$\mu$s.
This separation arises because switching a trap's occupancy is local and
immediate, whereas re-establishing the band-bending waits on the whole charge
environment to equilibrate.

Taken together, we expect the interface asymmetry to dominate the static field,
while both bulk and interface traps contribute to the noise.
The above-band pulse can thus be understood as an optical stand-in for surface
passivation, and what remains is a bulk residual: the slowly fluctuating
environment that persists once the interfaces are saturated and that limits the
coherent drive of \cref{fig:rabi}.

\section{Coherent control of reset emitters}
Having established methods to tune the T-center local field environment, we can
determine an optimal combination of $\Eext$ and $\sigma_E$ to demonstrate
improved T-center coherence.
We probe a spectrally isolated T-center at 1325.38~nm using fiber excitation and
collection to increase the excitation cross-section, and drive it with resonant
pulses of variable width (4--70~ns) after a free-space AB reset (Methods).

\begin{figure}[htbp]
    \centering
    \includegraphics[width=0.6\linewidth]{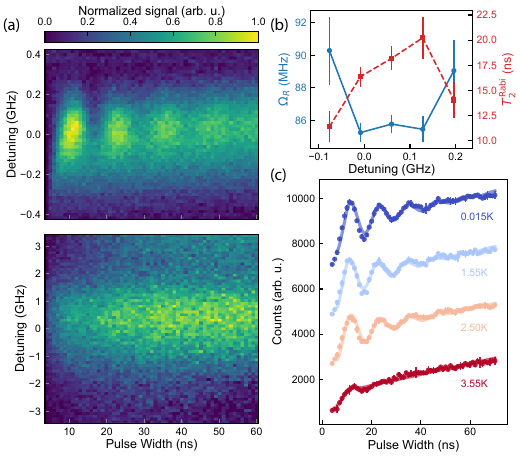}
    \caption{\textbf{Coherent optical control of a reset T-center} (a)
    Rabi pattern (laser frequency versus pulse width) of a spectrally isolated
    T-center with the inclusion (top) and absence (bottom) of an AB reset
    sequence; both background-subtracted for the pulse-width-dependent laser
    leakage of inline probing.
    (b) Fitted Rabi frequency and $T_2^{\mathrm{Rabi}}$ versus detuning; the weak frequency
    dependence indicates residual broadening.
    (c) Temperature-dependent Rabi oscillation at 60~MHz detuning: coherent
    oscillation extends to 2.5~K, and is heavily damped at 3.55~K.}
    \label{fig:rabi}
\end{figure}

The inclusion of an AB-pulse enables the coherent optical control of the T-center
(\cref{fig:rabi}a top).
The optical Rabi pattern (drive frequency versus pulse width) shows clear oscillation
and decay with the reset. In the absence of an optical reset, only a monotonic rise
is observed (\cref{fig:rabi}a bottom).

The Rabi frequency is nearly independent of detuning, shifting only from 85(1)
to 90(2)~MHz across 273~MHz (\cref{fig:rabi}b), deviating significantly from the expected
$\sqrt{\Omega_0^2 + \delta^2}$ of an ideal two-level system.
We consider this observation as a signature of residual spectral-diffusion
broadening--at each nominal detuning the laser addresses the sub-population
shifted into resonance by slow charge fluctuations that the reset does not fully
remove, and the coherence time $T_2^{\mathrm{Rabi}}$ falls as the detuning grows.
The longest coherence time we resolve is $T_2^{\mathrm{Rabi}} = 20(2)$~ns.

Finally, warming the device damps the oscillation while leaving the PLE
linewidth broadly unchanged (\cref{fig:rabi}c).
The coherent oscillation signal survives to 2.5~K, but is heavily damped by 3.55~K, as
thermal TX0--TX1 mixing increases the homogeneous linewidth~\cite{bergeron2020tcenter}.
The Rabi decay thus gives access to the homogeneous linewidth below the
sensitivity of PLE.

\section{Conclusions}

We have demonstrated multifaceted optical-electrical characterization, analysis,
and control over the local environment of a nanophotonic p--i--n-junction
integrated single T-center.
Implementing our findings, we observe unheralded optical linewidth narrowing
to a minimum of 128(22)~MHz (median--0.57(11)~GHz), the narrowest unheralded linewidth reported for an integrated T-center, with
statistics reported surveying 46 separate emitters.
Coupled with orthogonal electrical control over the second-order Stark effect,
we develop a model to interpret the proximal field as
possessing distinct components at equilibrium and of Gaussian stochastic noise.
An above-band pulse serves to reset the equilibrium field, and narrows the
noise profile.
Finally, we show that the optical reset suppresses the charge noise sufficiently to
permit all-optical coherent control of a narrowed T-center, with a $T_2^{\mathrm{Rabi}}$
coherence time of 20(2)~ns.

Beyond the linewidth improvement, these measurements use the T-center as a probe of its own charge environment, resolving trap dynamics having microsecond time scales, and revealing the field- and temperature-dependences that determine optical coherence.
While p--i--n integration and above-band control have each been demonstrated
separately in 4H-SiC, silicon, and quantum dots~\cite{anderson2019electrical,day2024electrical,majumdar2011photocarriers,pedersen2020transformlimited},
combining them in a single device renders the equilibrium and stochastic components of the environmental field separately addressable.
For frequency tuning, the correlation we observe between linewidth and
red-shift magnitude suggests strain may prove a better handle than electric field.

Although all currently reported integrated results remain three orders of magnitude from
the intrinsic lifetime-limited linewidth, our method already moves the emitter
toward a useful regime for cavity coupling.
The 20(2)~ns coherence time we resolve is within a factor of three of the
Purcell-enhanced lifetimes demonstrated for T-centers in photonic crystal
cavities, $T_{1,\mathrm{Purcell}} = 62(2)$--$363(56)$~ns~\cite{johnston2024cavity,komza2025multiplexed}.
The remaining requirement is the linewidth: recent theory projects that a 100~MHz
T-center in a 200k quality-factor cavity can provide readout fidelity
$F > 99.0\%$~\cite{wong2025readout}, a threshold the narrowest emitters reported
here can approach.

Our findings show that the noise environment can be dynamically engineered: the
AB-pulse optical reset saturates trap-induced noise sources.
Further linewidth narrowing will require materials engineering as well.
Two levers are paramount: improving the T-center formation yield, so that less
creation damage can be achieved, and post-fabrication surface treatment~\cite{islam2026passivation}.
This work provides both a deepened understanding of the underlying limitations
and a path toward their improvement to ultimately demonstrate a useful T-center
quantum node.

\section{Methods}

\subsection{Device design}

Devices are fabricated on high-resistivity silicon-on-insulator wafers (from SEH) with a
220~nm device layer. Ridge waveguides are defined by a 60\%
partial etch, leaving a slab that gives the buried junction
continuous field access to the emitters; the inverse-tapered fiber couplers
and Bragg back-reflectors that terminate each waveguide are fully etched
(\cref{fig:intro}a). The Bragg grating acts as a back-reflector to improve
collection. The inverse-tapered fiber couplers undergo a masked oxide release
process for collection efficiency improvement. A lateral p--i--n junction is
formed by implanting boron (30~keV) and phosphorus (75~keV) dopants, each at
$1.5\times10^{15}$~cm$^{-2}$, on either side of an intrinsic gap, targeting a
peak concentration of $2\times10^{20}$~cm$^{-3}$ at 110~nm depth. The
intrinsic-gap (junction) width is varied from 3 to 14~$\mu$m.

T-centers are created in the intrinsic region by sequential carbon and hydrogen
implantation, C at $1\times10^{12}$~cm$^{-2}$ (37~keV) and H at
$7\times10^{12}$~cm$^{-2}$ (9~keV).  The sample is annealed at 900~$^{\circ}$C for
3 mins after C implantation, and is followed by a rapid thermal anneal above
400~$^{\circ}$C after H implantation. The low carbon fluence is chosen to place
T-centers at a low areal density.

\subsection{Experimental setup}

The device is mounted in a dilution refrigerator (BlueFors LD250 series) at a base
temperature of 15~mK. Resonant excitation and the off-resonant
stabilization pulse are provided by two tunable diode lasers (Santec) in the
telecom O-band: the resonant probe is gated by a semiconductor
optical amplifier (SOA), and the 1325~nm stabilization pulse by an acousto-optic
modulator (AOM); the SOA and AOM provide 64 and 60~dB of pulse extinction
respectively. Above-band excitation is provided by a 640~nm pulsed diode laser
(Thorlabs NPL64B) co-propagating with the resonant beam.
Except for the Rabi measurement, all lights are delivered
in free space: a collimated beam is steered by a galvanometer and focused onto
the waveguide-coupled emitter through the cryostat top window and an 8-f optical
relay. Emission is collected into a lensed single-mode fiber; the resonant laser
and the above-band light are rejected by a 1350~nm longpass and a 1450~nm shortpass
filter pair, and photons are detected by a telecom-optimized
superconducting-nanowire single-photon detector (Single Quantum, 1325~nm
optimized). Junction bias is applied through a source-measure unit (Keithley) for
DC-biased PLE and an arbitrary-waveform generator (Siglent SDG series) for pulsed
biasing.

\subsection{Measurement modes}

All optical measurements are variants of pulsed photoluminescence excitation
(PLE): a resonant pulse near 1326~nm excites the T-center, and the resulting
emission is read out on the phonon sideband. The center
wavelength and the Gaussian-fit full width at half maximum (FWHM) are extracted from PLE brightness spectrum.
The baseline resonant probe is a 300~ns pulse at 100~$\mu$W at the device with a
900~ns collection window. Counts are normalized either by the number of
acquisition repetitions or by the maximum collected count, and unless stated
otherwise the background is subtracted after a gaussian or skew Voigt fit.

\bmhead{Above-band reset (AB-PLE)}
The AB reset is a train of 640~nm pulses from the co-propagating pulsed diode
laser, each 39~ns long at 0.33~mW (unattenuated) at the device, followed by the
resonant PLE probe after a delay of 3--10~$\mu$s unless otherwise stated. Three
parameters are swept to characterize and optimize the reset (\cref{fig:absweep}): the
AB-to-probe delay, the pulse power set by a free-space attenuator, and the
number of pulses per shot. AB-PLE spectra sit on an elevated background, which
is subtracted after fit.

\bmhead{Automated emitter screening}
The chip-scale statistics of \cref{fig:stats} come from an automated screen run along each
waveguide. A coarse inline PLE scan over 1325.3--1326.2~nm lists ten to fifteen
candidate wavelengths per waveguide; the excitation spot is then raster-scanned
at each to retain only spatially isolated spots; and the laser position is
optimized twice against CW-PLE counts, with a fine PLE scan setting the center
wavelength. An SNR threshold is applied to ensure that every retained emitter
supports a reliable lineshape fit.
Each qualified emitter is measured as a matched PLE / AB-PLE pair,
and the pipeline fits both lines for Gaussian FWHM, center wavelength, and
AB-to-PLE shift. The screen is blind to both extremes: lines too broad fall below
the single-scan SNR threshold, and lines too narrow can be stepped over in the
spatial raster.

\bmhead{AWG-synchronized pulsed biasing}
To Stark-tune the emitter while holding its charge environment stable, the
junction is driven by an arbitrary-waveform generator (AWG): a square wave (\cref{fig:field}a inset) at $\sim$30\% duty cycle whose high level
holds the tuning bias $V$ during the resonant PLE probe and whose low level
returns to 0~V. An off-resonant 1325.0~nm stabilization pulse (the ``IR
stabilization pulse''), of approximately the probe power and $\sim$10$\times$ its
duration, is triggered toward the middle of the low half and re-equilibrates the
trapped-charge environment between probes. Replacing this IR pulse with an AB
reset pulse gives the AB-anchored biasing of \cref{fig:field}f.
The stabilization pulse is adopted empirically. Without it, repeated bias sweeps left the ZPL shifting more abruptly than the quadratic Stark response and dimming at lower bias; interleaving the off-resonant pulse suppressed both. We attribute the rapid drift and dimming to a light-induced space-charge distribution that accumulates under resonant probing at high field, leaving a memory dependent local-field; the below-bandgap pulse redistributes that charge between probes, so each sweep starts from the same configuration. We therefore refer to it as a stabilization pulse.

\bmhead{Ramped-bias field ionization}
To probe field ionization at high bias (\cref{fig:field}e), the junction bias is ramped
over 500~ns and the excited-state lifetime is read from the photon-arrival
histogram.

\bmhead{Pseudo-CW PLE (time-resolved recovery)}
To resolve how the charge environment recovers after a reset (\cref{fig:dynamics}), five AB
pulses reset the environment and a single 20~$\mu$s resonant pulse at
100~$\mu$W then probes it; binning the photons by arrival time within that
window gives a time-resolved spectrogram (SNR $>$5 for TC3). The scheme works
because the resonant signal-to-noise ratio exceeds unity within one pulse.

\bmhead{Coherent optical control (Rabi)}
For coherent driving (\cref{fig:rabi}) the same T-center is addressed with fiber
excitation \emph{and} collection to increase the excitation cross-section,
following a free-space AB reset. Resonant pulses of variable width (4 to 70~ns)
drive the TX0 transition at 1325.38~nm, and scanning drive frequency against
pulse width maps the Rabi pattern; traces are corrected for the
pulse-width-dependent laser leakage of inline probing.

\subsection{Analysis model}

\bmhead{The model}
A T-center's transition frequency responds to the net electric field at the
emitter through a quadratic Stark map,
\begin{equation}
  \Delta\nu(\Eeff) = A\,\Eeff + \alpha\,\Eeff^{2},
  \label{eq:stark}
\end{equation}
with $\alpha = 0.115$~Hz\,m$^{2}$/V$^{2}$ and $|A| = 8335$~Hz\,m/V taken from
Clear2024 rather than fitted~\cite{clear2024transition}. We split the net field into a quasi-static part and
Gaussian charge noise,
\begin{equation}
  \Eeff = \Eext + \mathcal{N}(0, \sigma_E),
  \label{eq:field-split}
\end{equation}
so that two numbers describe an emitter's electrostatic environment: $\Eext$,
referenced to the AB line as zero deterministic field, and the charge-noise
amplitude $\sigma_E$. These two are the only quantities the model returns.

\bmhead{Two free parameters to set lineshape}
Averaging the homogeneous line $L$, a Lorentzian of fixed FWHM
$\Gamma_{\mathrm{hom}} = 0.10$~GHz, over the noise distribution $\Phi$ of
\eqref{eq:field-split} gives the observed lineshape,
\begin{equation}
  I(\nu) = \int\!\dE\; \Phi(E_{ce})\,
    L\!\left(\Delta\nu(\Eext + E_{ce}) - \nu,\ \Gamma_{\mathrm{hom}}\right),
  \label{eq:lineshape}
\end{equation}
which has no free parameters for its center and width, since $(\Eext, \sigma_E)$ fixes all
three of its features:
\begin{center}
\begin{tabular}{ll}
  \toprule
  feature & fixed by \\
  \midrule
  position  & $A\Eext + \alpha\Eext^{2}$, plus a rectification
              $\alpha\sigma_E^{2}$ from the curvature \\
  width     & $\sigma_\nu^{2} = g^{2}\sigma_E^{2} + 2\alpha^{2}\sigma_E^{4}$,
              with $g = A + 2\alpha\Eext$, measured as
              $\Gamma_{\mathrm{SD}} = 2\sqrt{2\ln 2}\,\sigma_\nu$ \\
  asymmetry & the curvature of \eqref{eq:stark} \\
  \bottomrule
\end{tabular}
\end{center}
Fitting a spectrum to \eqref{eq:lineshape} returns both numbers directly, and
every model quantity plotted in the figures is one of the two.
$\Gamma_{\mathrm{hom}} = 0.10$~GHz is not fitted but taken as a conservative
ceiling, so the extracted $\sigma_E$ is a lower bound. The coherent driving of
\cref{fig:rabi}a places the true value well below it: $T_2^{\mathrm{rabi}} = 20(2)$~ns
bounds $\Gamma_{\mathrm{hom}} \leq 17$~MHz, consistent with hole-burning
measurements in comparable devices~\cite{bowness2025spectral,islam2026passivation}.
Varying $\Gamma_{\mathrm{hom}}$ over that range changes $\sigma_E$ by less than
8\% and leaves every comparison reported here unchanged.

\bmhead{Guardrails to the analysis}
Uncertainties in parentheses are statistical, from the fits. Derived quantities
for which calibration systematics dominate are quoted without uncertainties.
Absolute $\sigma_E$ and $\Eext$ should be read to within a factor of two. The Stark
coefficients are transferred from bulk $^{28}$Si to natural-abundance SOI, and
their vector and tensor forms are reduced to the scalars $A$ and $\alpha$ for
tractability, so an emitter's orientation relative to the applied field is not
resolved.
Comparisons across
emitters, biases and delays share the same coefficients and are unaffected. The
dynamics rest on a single emitter, and the sheet geometry idealizes the
device as two parallel interfaces, neglecting the etched ridge sidewalls.

\section*{Data availability}
The data that support the findings of this study are available from the
corresponding author upon request.

\section*{Note added}

During manuscript preparation, we became aware of recent complementary manuscripts
presenting optical linewidth narrowing using surface passivation and above-band
illumination \cite{islam2026passivation,johnston2026}.

\section*{Acknowledgements}

The authors thank Chang Jin, Amberly Xie, Guanhao Huang, Di Liu, Jonathan Dietz,
Matthew Yeh, Neil Sinclair and Zihuai Zhang for helpful discussions on
experimental details.
Anthropic generative AI tools were used as a supplemental research aid for instrument control,
data fitting and visualization, and language polishing; all outputs were
verified by the authors. Simulations related to the photonics design were
performed under support of Tidy3D education license.

This work was supported by Amazon Web Services (AWS) Awards A50791 and A60290
and the Harvard Quantum Initiative. Additional support was provided by the U.S. Department of Energy, Office of Science, Office of Basic Energy Sciences, Materials Sciences and Engineering Division under Contract No. DE-AC02-05-CH11231 within the Quantum Coherent Systems Program (KCAS26) for ultralow temperature optical measurements and by the US Department of Energy, Office of Science, Basic Energy Sciences in Quantum Information Science under Award No. DE-SC0022289 for qubit synthesis and characterization.

The devices used in this work were fabricated at the Berkeley Marvell NanoLab
and at the Harvard University Center for Nanoscale Systems (CNS), a member of
the National Nanotechnology Coordinated Infrastructure Network (NNCI), which is
supported by the National Science Foundation under NSF award no. ECCS-2025158.

\section*{Author contributions}

Methodology: C.Z., H.S., L.K., A.S., E.L.H.; Fabrication: C.Z., H.S., L.K.,
D.W.; Measurement: C.Z., H.S., E.G.; Analysis: C.Z., A.M.D., E.G., A.S.,
E.L.H.; Advising: M.K.B., A.S., E.L.H.; Manuscript preparation: All authors.



\bibliographystyle{unsrt}
\bibliography{refs}

\end{document}